\documentclass[useAMS,usenatbib]{mn2e}
\usepackage{natbib}
\usepackage{amssymb,amsmath,amsfonts}
\usepackage{epsfig}

\def\reff@jnl#1{{\rm#1\/}}

\def\aj{\reff@jnl{AJ}}                  
\def\araa{\reff@jnl{ARA\&A}}            
\def\apj{\reff@jnl{ApJ}}                        
\def\apjl{\reff@jnl{ApJ}}               
\def\apjs{\reff@jnl{ApJS}}              
\def\ao{\reff@jnl{Appl.Optics}}         
\def\apss{\reff@jnl{Ap\&SS}}            
\def\aap{\reff@jnl{A\&A}}               
\def\aapr{\reff@jnl{A\&A~Rev.}}         
\def\aaps{\reff@jnl{A\&AS}}             
\def\azh{\reff@jnl{AZh}}                        
\def\baas{\reff@jnl{BAAS}}              
\def\jrasc{\reff@jnl{JRASC}}            
\def\memras{\reff@jnl{MmRAS}}           
\def\mnras{\reff@jnl{MNRAS}}            
\def\pra{\reff@jnl{Phys.Rev.A}}         
\def\prb{\reff@jnl{Phys.Rev.B}}         
\def\prc{\reff@jnl{Phys.Rev.C}}         
\def\prd{\reff@jnl{Phys.Rev.D}}         
\def\prl{\reff@jnl{Phys.Rev.Lett}}      
\def\pasp{\reff@jnl{PASP}}              
\def\pasj{\reff@jnl{PASJ}}              
\def\qjras{\reff@jnl{QJRAS}}            
\def\skytel{\reff@jnl{S\&T}}            
\def\solphys{\reff@jnl{Solar~Phys.}}    
\def\sovast{\reff@jnl{Soviet~Ast.}}     
\def\ssr{\reff@jnl{Space~Sci.Rev.}}     
\def\zap{\reff@jnl{ZAp}}                        
\def\nat{\reff@jnl{Nature}}             
\def\physrep{\reff@jnl{Phys.~Rep.}}    

\title[Pairwise velocities in the Halo Model: Luminosity and Scale Dependence]{Pairwise velocities in the Halo Model: Luminosity and Scale Dependence}
\label{firstpage}
\author[Slosar et al]  {An\v{z}e Slosar$^1$, Uro\v{s}
  Seljak$^{2,3}$, Argyro Tasitsiomi$^4$\\
  $^1$Faculty of Mathematics and Physics, University of Ljubljana,
  Slovenia \\
  $^2$International Centre for Theoretical Physics, Trieste, Italy \\
  $^3$Department of Physics, Princeton University, Princeton, NJ
  08544, USA \\
  $^4$Department of Astronomy \& Astrophysics, Kavli Institute for
  Cosmological Physics, The University of Chicago, Chicago, IL 60637, USA
}

  \date{Accepted ---; received ---; in original form \today}
\pagerange{\pageref{firstpage}--\pageref{lastpage}}
\pubyear{2002}

\newcommand {\rmd}{{\rm d}}
\newcommand {\rmdm}{{\rm dm}}

\newcommand {\rmc}{{\rm C}}
 
\newcommand {\rmnc}{{\rm NC}}

\newcommand {\rmhh}{{\rm hh}}
\newcommand {\rmhs}{{\rm hs}}
\newcommand {\rmss}{{\rm ss}}

\newcommand {\sinc}{\operatorname{sinc}}
\newcommand {\rmlin}{{\rm lin}}

\newcommand {\proj}{{\rm proj}}
\newcommand {\cyl}{{\rm cyl}}

\newcommand{\mpch}{{h^{-1}\ \rm Mpc}}

\newcommand{\hmodot}{{h^{-1} M_\odot}}

\newcommand{\vect}[1]{\mathbf{#1}}

\begin{document}
\maketitle

\begin{abstract}
  We investigate the properties of the pairwise velocity dispersion as
  a function of galaxy luminosity in the context of a halo model. We
  derive the distribution of velocities of pairs at a given separation
  taking into account both one-halo and two-halo contributions. We
  show that pairwise velocity distribution in real space is a
  complicated mixture of host-satellite, satellite-satellite and
  two-halo pairs. The peak value is reached at around 1$h^{-1}$Mpc and
  does not reflect the velocity dispersion of a typical halo hosting
  these galaxies, but is instead dominated by the satellite-satellite
  pairs in high mass clusters. This is true even for
  cross-correlations between bins separated in luminosity.  As a
  consequence the velocity dispersion at a given separation can
  decrease with luminosity, even if the underlying typical halo host
  mass is increasing, in agreement with recent observations.  We
  compare our findings to numerical simulations and find a good
  agreement.  Numerical simulations also suggest a luminosity
  dependent velocity bias, which depends on the subhalo mass.  We
  develop models of the auto- and cross-correlation function of
  luminosity subsamples of galaxies in the observable $r_\proj - \pi$
  space and calculate the inferred velocity dispersion as a function
  of wave vector if dispersion model is fit to the redshift space power
  spectrum. We find that so derived pairwise velocity dispersion also
  exhibits a bump at $k\sim 1 h/{\rm Mpc}$.
\end{abstract}


\section{Introduction}

Large redshift surveys such as SDSS \citep{2000AJ....120.1579Y} and
2dF \citep{2001MNRAS.327.1297P} have provided us with a unique tool to
probe the 3-dimensional structure of our Universe. The radial
distances in those surveys are, however, modulated by the components
of peculiar velocities along the line of sight.  On large scales these
velocities can be modelled using linear theory, as developed by
\cite{1987MNRAS.227....1K}. On small scales the velocities lead to
radial stretching of groups and clusters, the so-called
``fingers-of-God''(FOG). In principle this anisotropy in the observed
structure can give us information on velocity structure of the
Universe in addition to the spatial distribution of objects. Peculiar
velocities of individual objects are sensitive to the amount and
distribution of dark matter, but are difficult to measure in the
absence of reliable distance estimators.  The projected pairwise
velocity differences, on the other hand, are much easier to measure
from those surveys.  Understanding the pairwise velocity distribution
on all scales could therefore provide important dynamical information
on the relation between the galaxies and surrounding dark matter.

The simplest approach to use dynamical information from pairwise
velocities is to select isolated halo systems by some scheme
\citep{2002ApJ...571L..85M,2003ApJ...598..260P,2003ApJ...593L...7B,2004MNRAS.352.1302V}.
This process usually results in a selection of several system composed
of a host galaxy and their satellites. Models can then be tested when
many such isolated systems are compared ``on average'' to the
theoretical predictions. The main advantage of this approach is that
the selection scheme already labels the host and satellite galaxies,
making the modelling somewhat easier. However, this approach also has
several disadvantages: the number of suitable galaxies is very small
(of the order of 1\% of the total number of galaxies), the treatment
of interlopers (galaxies that appear to be part of the system due to
projection effects) can be difficult and model dependent, some of host
galaxies may themselves be satellites of a larger system and, lastly,
selecting isolated galaxy-satellite systems may not be representative
of the dark matter-galaxy relation in general.

The alternative approach is to do as little preselection as possible
and instead model the effects statistically within a suitable model.
This has the advantage of having large statistics and being more
representative, given that by definition all galaxies are being used.
This is the approach taken in this paper.  We focus on the small scale
correlations in redshift space and use the halo model \citep[see
e.g.,][]{2002PhR...372....1C} to provide an interpretation of the
observations in the context of a physical model.  Our approach thus
focuses on statistical properties of the entire distribution of
galaxies.  Since the standard host-satellite separation is based on
luminosity, this suggests that cross-correlation analysis between
faint and bright galaxies might be especially effective in selecting
central-satellite pairs and providing information about the halo
structure.  We thus follow \cite{2002MNRAS.335..311G} in working with
narrow luminosity bins and perform both auto and cross-correlation
analysis between them.

We assume the galaxies are either central 
galaxies in a halo or satellites, both of which need to have a specified
conditional halo mass probability distribution.  In this model the
contributions to pairwise velocities come from host-satellite
(host-sat from now on) and satellite-satellite (sat-sat from now on)
pairs within the same halo and from two halo pairs (both host and
satellite).  A related  approach with luminosity thresholds has been
developed by \cite{2004astro.ph..8564Z}, while \cite{2004MNRAS.352.1302V} have focused on conditional
luminosity function as a function of halo mass.  We first build a
model to describe the velocity structure of galaxies in the dark matter
halos (Section \ref{sec:pairw-veloc-halo}). We perform this in terms of
velocity distribution function, the probability that a pair of
galaxies belonging to a given luminosity bin can be found at a given
separation and has a given velocity component.  This procedure is
in spirit similar to that of \cite{1996MNRAS.279.1310S}, although with
a different emphasis and a more up to date version of the halo
model. We show that less luminous galaxies can appear to move faster
than their typical parent halo population even in the simplest halo
model, in agreement with some observational results
\citep{2004ApJ...617..782J}.

In Section \ref{sec:basic-pred-comp} we compare our model with
numerical simulation.  We calculate the correlation function in the
$r_\proj-\pi$ plane and use it to predict the dependence of
$\sigma_{12}$ parameter one would obtain by fitting it with a simple
dispersion model (Section \ref{sec:distr-r_pr-pi}). The last section
summarises our findings and concludes the paper.  Throughout this
paper, we assume a concordant flat cosmology with $\Omega_m=0.3$,
Hubble constant of $70 {\rm km\ s}^{-1}{\rm Mpc}^{-1}$ and
$\sigma_8=0.9$.

\section{Pairwise velocities in halo model}
\label{sec:pairw-veloc-halo}

We follow the standard halo model, which postulates
that all galaxies live in the dark matter halos. The mass function $\eta =
\rmd n/\rmd M$ of dark matter halos is most reliably estimated from
numerical simulations and here we use the mass function fit from
simulations \citep{2001MNRAS.321..372J}
\begin{equation}
 \eta= \frac{\rmd n}{\rmd M} = 0.315 \frac{\rho_0}{M^2} \exp \left[ - | \ln
  \sigma^{-1} + 0.61 | ^ {3.8} \right ] \frac{\rmd \ln
  \sigma^{-1}}{\rmd \log M},  
\end{equation}
where $\sigma$ is the variance in the linear density field at a given
redshift, after smoothing with a spherical top-hat filter which
encloses mass $M$ in the mean. Each halo has a central galaxy and a
Poisson distributed number of satellite galaxies. We consider galaxies
with r-band luminosity between $L=-19$ and $L=-22$ and split them into
three luminosity bins of unit magnitude size. We always refer to a
luminosity bin by its faint end magnitude. Additionally, we will refer
to a sample of galaxies selected so that each pair contains one galaxy
from luminosity bin $L_i$ and one galaxy from the luminosity bin $L_j$
as ($L_i$, $L_j$). We will use symbols $n_\rmc(M|L_i)$ and
$n_\rmnc(M|L_i)$ to denote the expectation value of the number of
galaxies per halo which belong to the luminosity bin $L_i$ and occupy
central and non-central positions, respectively.  Since every halo is
assumed to have a central galaxy, the following must be true:
\begin{equation}
  \sum_i n_\rmc(M|L_i) = 1,
\end{equation}
although some of these galaxies will fall outside the luminosity 
range we consider here. 

We assume that the central galaxy is at the centre and at rest with
respect to the parent halo. The satellite galaxies are distributed
in the parent halo around the central galaxy. We assume this radial distribution 
follows the NFW profile
\citep{1996ApJ...462..563N}.  We denote the probability of finding a galaxy at a
distance between $r$ and $r+\rmd r$ in a halo of mass $M$ as $\Phi_1$:
\begin{equation}
  \Phi_1(r|M) \propto \frac{r}{r_s} \left(1+\frac{r}{r_s}\right)^{-2}.
\end{equation}
where $r_s = r_{\rm vir}/c$, where $c$ is concentration and $r_{\rm
  vir}(M)$ is virial radius, which is defined as a radius of a sphere
of mass $M$ which has the average density 200 times critical density
of the Universe. In simulations the subhalo distribution is less
concentrated than the dark matter. Whether the same is true in the
real universe is still a matter of an ongoing debate
\citep{2000ApJ...544..616G,2001astro.ph..9001B,2004MNRAS.352..535D,2005MNRAS.361..415C,2005MNRAS.362..711Y,2005ApJ...618..557N}.
In this work we assume the concentration parameter for the satellite
galaxy distribution $c_g=3$ and a cut-off at the $r_{\rm
  cut}=1.5r_{\rm vir}$, which is in a reasonable agreement with
simulations. Simulations show a weak dependence of the best-fit
concentration on luminosity, but we neglect this effect here.

The probability of finding a galaxy belonging to the luminosity bin
$L_i$ in a halo with mass between $M$ and $M+\rmd M$ can be split into
contribution from central and satellite galaxies:
\begin{equation}
  P(M|L_i) \rmd M = (1-\alpha) P_\rmc (M|L_i) + \alpha P_\rmnc (M|L_i),
\end{equation}
where $\alpha$ is the fraction of satellites among galaxies 
of luminosity $L_i$. 

The probability distribution for the central galaxy is often assumed
to be a $\delta$ function. Here we assume a slightly more realistic model
and introduce a Gaussian spread around the mean value of $\log M$
(hereafter we assume $\log$ to be base 10). Although real
distributions are asymmetric, especially at higher luminosities, it is
a useful first-order correction. Therefore,
\begin{equation}
  P_{\rmc,i} (M) = G(\log M; \log M_{0,i}, \sigma_{v,i}),
\end{equation}
where $G(x; \mu, \sigma)$ denotes a Gaussian distribution with mean
$\mu$ and dispersion $\sigma$.

Following \cite{2004astro.ph.10711M} the 
distribution for $P_{\rmnc,i}$  is assumed to reflect a double power
law for the number of satellites per halo:
\begin{equation}
  P_{\rmnc,i} (M) = \begin{cases}
    0 & M< M_{0,i} \\ 
    \propto \eta M^2  & M_{0,i} < M < 3 M_{0,i} \\
    \propto  \eta M & \mbox{otherwise}\\
  \end{cases}
\label{nm}
\end{equation}
Constants of proportionality are chosen so that the function is
continuous and normalised and give the overall satellite fraction 
$\alpha_i=0.2$.
Based on simulations 
we choose the following canonical values for $M_{0,i}$: $M(-19) = 5
\times 10^{11} \hmodot$, $M(-20) = 2 \times 10^{12} \hmodot$, $M(-21)
= 1 \times 10^{13} \hmodot$ and fix $\sigma_{v,i}=0.2$.

The total number of galaxies is given by the product of the number of
halos and the number of galaxies per halo and therefore
\begin{eqnarray}
P_\rmc (M|L_i) = \frac{\eta n_\rmc(M|L_i)}{\int \eta n_\rmc(M|L_i) \rmd M} =
\frac{\eta n_c(M|L_i)}{N_{\rmc,i}},\label{eq:3}\\
P_\rmnc (M|L_i) = \frac{\eta n_\rmnc(M|L_i)}{\int \eta n_\rmnc(M|L_i) \rmd M} =
\frac{\eta n_c(M|L_i)}{N_{\rmnc,i}},\label{eq:4}\\
\end{eqnarray}
where $N_{\rmc,i}$ and $N_{\rmnc,i}$ denote the total number density
of galaxies belonging to a central/non-central class and belonging to
the luminosity bin $L_i$. The fraction of non-central galaxies
$\alpha_i$ is therefore given by:
\begin{equation}
  \alpha_i = \frac{N_{\rmnc,i}}{N_{\rmnc,i}+N_{\rmc,i}}= \frac{N_{\rmnc,i}}{N_{i}}.
\end{equation}

Next we turn to the distribution of velocities. Let us then define the
pairwise velocity distribution $P(v,r|L_i,L_j)$ as the
probability that of all pairs between galaxies from luminosity bin
$L_i$ and galaxies from the luminosity bin $L_j$, we find one at
separation $r$, whose velocity difference
vectors has magnitude $v$. We emphasise that in our definition, this
is an object-weighted quantity rather than volume weighted
quantity. The pairwise velocity dispersion as a function of radius is
then given by:
\begin{equation}
  \sigma_2^2 (r|L_i,L_j) = \frac{\int_{0}^{\infty}
  P(v,r | L_i, L_j) v^2 \rmd v} {\int_{0}^{\infty}  P(v,r |  L_i, L_j) \rmd v}.
\label{eq:10}
\end{equation}
Similarly the correlation function is given by:

\begin{equation}
  1+\xi(r|L_i,L_j) = \frac{\int_{0}^{\infty}  P(v,r | L_i, L_j) \rmd v}{4 \pi r^2 N_i N_j}
\label{eq:7}
\end{equation}

The pairwise velocity has two contributions. Single halo contribution
comes from pairs residing in a single halo and dominates at small
distances ($\lesssim 1 \mpch$), while the two halo contribution dominates at larger
distances.

Therefore we can write
\begin{equation}
  P(v,r, |L_i, L_j) = P^{1h}(v,r|L_i,|L,j)+P^{2h}(v,r|L_i,L_j).
\end{equation}
We now investigate each contribution in turn.


\begin{figure}
  \epsfig{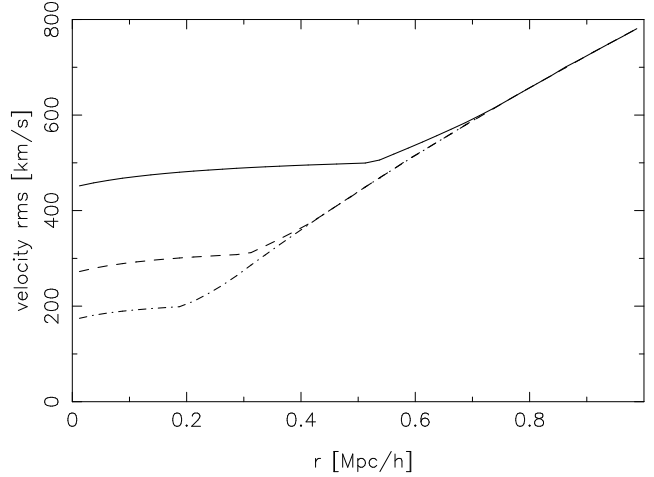}
  \caption{This figure shows the predictions for the one-point
  velocity dispersion relative to the centre of the halo for
  satellites selected from different luminosity bins for the
  halo model. Three luminosity bins are plotted: $-21$ (solid
  line), $-20$ (dashed line), $-19$ (dot dashed line). \label{fig:pp1}}
\end{figure}


\subsection{One-halo contribution}

The single halo contribution to the pairwise velocity distribution has
contributions from pairs formed one from the central and one from the satellite
galaxies (host-sat pairs, denoted as hs) and those coming from two satellite galaxies
(sat-sat pairs, denoted as ss):
\begin{equation}
  P^{1h}(v,r| L_i,L_j) =  P_\rmhs(v,r| L_i,L_j) + P_\rmss(v,r| L_i,L_j).
\end{equation}

The host-sat contribution is proportional to
\begin{multline}
  P_\rmhs (v,r|L_i,L_j) \propto \int   \big[
 n_\rmc(M|L_i)n_\rmnc(M|L_j) + \\
 n_\rmc(M|L_j)n_\rmnc(M|L_i) \big] P_{1v}(v|r,M)  
\Phi_1(r|M) \eta \rmd M.
\label{eq:1}
\end{multline}

Note that even though formally we have two terms inside 
square brackets of equation \ref{eq:1}, if $L_i<L_j$
first term is negligible by equation \ref{nm}. 

The 1-particle velocity distribution $P_{1v}(v|r,M)$ is assumed to be
Maxwellian. This is true for isothermal spheres and is in reasonable
agreement with simulations. Therefore
\begin{equation}
  P_{1v}(v|r,L_i) = M(v; b_{\rm v}(L_i)\sigma_{\rmdm}),
\label{eq:8}
\end{equation}
where $M(x; \sigma)$ denotes a Maxwell distribution corresponding to
one-dimensional velocity variance $\sigma_{\rmdm}$, which we take to be
\begin{equation}
     \sigma_{\rmdm} = 
90 {\rm km s^{-1}}
\left( \frac{M}{10^{12} \hmodot} \right)^{1/3}. 
\end{equation}
For the sake of simplicity we neglect the velocity bias at this stage,
simply setting $b_{\rm v}=1$, but we return to this point below when we
discuss the simulations.  Note that we can simply switch between
calculating the distribution of one-dimensional components of velocity
and the distribution of speeds (lengths of the 3-dimensional velocity
vectors) by replacing the Maxwell distribution with the corresponding
one-dimensional Gaussian. We will use the superscript ${\rm 1D}$ to
denote the distribution of one-dimensional components of velocity.

Similarly, one can write the sat-sat contribution as:
\begin{multline}
  P_\rmss (v,r|L_i,L_j) \propto \int   
 n_\rmnc(M|L_i)n_\rmnc(M|L_j) \\
\times P_{2v}(v|M)  \Phi_2(r|M) \eta \rmd M.
\label{eq:2}
\end{multline}
In this equation $\Phi_2$ denotes the probability that a pair of
satellites is found at separation $r$. This corresponds to the
self-convolution of the profile, which is most easily achieved in the
Fourier space:
\begin{equation}
  \Phi_2(r|M)  \propto r^2 \int \left[ \int \rmd r {\rm \Phi_1(r|M)} \sinc(kr) \right]^2 \sinc(kr) k^2 \rmd k.
\end{equation}
It should be emphasised that the halo occupations are stochastic and a
probability distribution is required to fully describe their
statistical properties.  We have implicitly performed the appropriate
averages in the above expressions. In particular, the number of pairs
for a Poisson distributed variable $p$ is given by $\left< p (p-1)
\right>=\left<p \right>^2$.



The function $P_{2v}(v|M)$ is used to denote the probability that the
difference between velocity vectors has magnitude $v$. Because we
assumed Maxwell distribution of velocities this is given simply by
\begin{equation}
  P_{2v}(v|r,L_i) = M(v; \sqrt{2} \sigma_{\rmdm}).
\end{equation}

The above equations can be combined in the following expression for
the pairwise velocity distribution:
\begin{multline}
  P^{1h}(v,r| L_i,L_j) =  \\
\int \bigg( \big[ (1-\alpha_i)\alpha_j
  P_{\rmc}(M|L_i)   P_{\rmnc}(M|L_j) P_{1v}(v|L_j)+ \\
(1-\alpha_j)\alpha_i P_{\rmc}(M|L_j)  P_{\rmnc}(M|L_i) P_{1v}(v|L_i)
  \big]
  \Phi_1(r|M)  + \\
 \alpha_i \alpha_j P_{\rmnc}(M|L_i)  P_{\rmnc}(M|L_j) \Phi_2(r|M)
  P_{2v}(v|L_i,L_j)\bigg) \\
 \times  N_i N_j \frac{\rmd M}{\eta}.
\label{eq:5}
\end{multline}
This can be meaningfully compared to the one-point velocity
distribution of galaxies at a distance $r$ from the centre of its
halo, that belong to the luminosity bin $L_i$.  In our notation, this is given by:
\begin{equation}
  P_1(v,r|L_i) = \int \alpha_i P_\rmnc(M|L_i) \Phi_1(r|M)
  P_{1v}(v|L_i) N_i \rmd M.
\label{eq:6}
\end{equation}
Note that the contribution due to the peculiar motion of the entire
halo is not present in the above distribution.  

Equations (\ref{eq:5}) and (\ref{eq:6}) show that true pairwise
velocities tend to favour higher velocities compared to the 
one-point velocity distribution for two reasons. First,
in a halo of a given mass, the velocity difference vectors of sat-sat
pairs have a higher dispersion (by a factor of $\sqrt{2}$ if we neglect
velocity bias). Second, the number of pairs contributed by a single halo
scales as the square of the number of its satellites and since heavier
halos contain more satellites, the pairwise velocity dispersion puts
more weight towards heavier halos.  This is encoded in the weighting
factor for true pairwise velocity distribution, which is $\eta^{-1}
\rmd M$ compared to $\rmd M$ for the case of one-point velocity
distribution.

\subsection{Two-halo contribution}

Two halo contribution comes from galaxies that are positioned in
different halos. For host-host pairs, the correlation function
follows the correlation function of parent halos, while for host-sat
pairs the correlation function would have to be convolved with the
halo profile and for the sat-sat it would have to be convolved with
the halo profile twice. However, these effects are only important 
on small scales where one-halo term dominates. Hence, 
we ignore this effect and assume that to a
good approximation the three correlation functions are equivalent.

The correlation function for halos follows that of the linear theory,
with the complication that the more massive halos are more
correlated. This is modelled in a simple manner by introducing a bias
factor for halos of mass $M$, given by \cite{1999MNRAS.308..119S}
\begin{equation}
  b(\nu) =  1 + \frac{\nu - 1}{\delta_c} + \frac{2p}{\delta_c(1+\nu'^p)},
\end{equation}
where $\nu = [\delta_c/\sigma(M)]^2$ and $\nu'=a\nu$. The spherical
over-density at which a clump collapses $\delta_c$ is $\delta_c=1.68$ for
Einstein-de Sitter model and fitted values of other parameters are
$a=0.73$ and $p=0.15$ \citep{2004astro.ph.10711M}. The cross-correlation between haloes of
different masses is then given by:
\begin{equation}
  \xi_{12}(r|M_1,M_2) = \xi_\rmlin(r) b(M_1) b(M_2).
\end{equation}

The probability of finding a pair of halos of masses between $M_{1,2}$
and $M_{1,2}+\rmd M_{1,2}$, separated by a distance between $r$ and
$r+\rmd r$ is therefore given by 
\begin{equation}
\left(1+\xi_{12}(r|M1,M2)\right) \eta(M_1) \eta(M_2)  4 \pi r^2.
  \end{equation}

Differences between host-host, host-sat and sat-sat pairs arise when
considering their velocities. The velocity difference between
host-host galaxy pairs arises purely due to velocity difference of
their parent halos. The velocity distribution for halos is often
assumed Maxwellian \citep{2001MNRAS.322..901S}. In our simulations, it
was observed that to a good approximation the velocity dispersion is
given by
$  \sigma_{\rm halo} = 455 {\rm km\ s^{-1}} $,
independent of halo mass.


\begin{figure}
  \epsfig{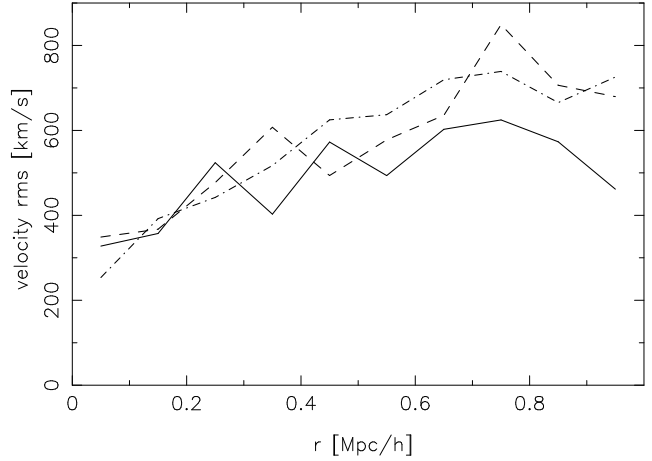}
  \caption{This figure shows the results for the one-point velocity
  dispersion relative to the centre of the halo in the
  simulation. Lines are labelled as in Figure
  \ref{fig:pp1}.\label{fig:spp1}}
\end{figure}


However, there are two factors that have to be taken into account when
calculating the two-halo velocity dispersion term. First,
neighbouring velocities are heavily correlated and consequently the
probability distributions for velocities of two halos cannot be
assumed independent. Second, if we select halos with a nearby
neighbour and compare its velocity dispersion to population of halos
without a nearby neighbour we find that the former is higher. This is
simply due to the fact that halos that have close neighbours are more
likely to live in a denser environment and thus have higher
velocity dispersion. These effects are difficult to model and we combine them
into an effective dispersion suppression factor $s(r)$, so that the
halo-halo probability for velocity difference can be written as:

\begin{equation}
  p^{2h}_{\rmhh} (v|r,M_1,M_2) = M(v; s(r)\sqrt{2} \sigma_{\rm halo})
\end{equation}
We use the following fitting formula for $s(r)$:
\begin{equation}
  s(r) = \begin{cases} 
  \left(\frac{r}{r_0}\right)^\nu & r<r_0 \\
  1 & {\rm otherwise}, \\
  \end{cases}
\end{equation}
where $r_0=50 \mpch$ and $\nu=0.15$. The relative bias of haloes of
different mass is ignored in this approximation.


When  considering host-sat and sat-sat pairs we have to add the
dispersion due to the satellite's peculiar motion with respect to its
parent halo centre:
\begin{equation}
  p^{2h}_{\rmhs} (v|r,M_1,M_2) = M\left(v; \left[ 2\sigma_{\rm
  halo}^2 s(r)^2 +\sigma_\rmdm(M_2)\right]^{1/2}\right)
\end{equation}

and
\begin{multline}
  p^{2h}_{\rmss} (v|r,M_1,M_2) = M \big(v; [ 2\sigma_{\rm
  halo}^2 s(r)^2\\
+\sigma_\rmdm(M_1)+\sigma_\rmdm(M_2)]^{1/2}\big).
\end{multline}

Finally, we can assemble the 2-halo probability function:
\begin{multline}
  P^{2h}(v,r|L_i,L_j) = \int \rmd M_1 \int \rmd M_2  \\
\times \bigg((1-\alpha_i)(1-\alpha_j)P_\rmc(M_1|L_i) P_\rmc(M_2|L_j)  p^{2h}_{hh}(v|M_1,M_2) \\
+P_\rmnc(M_1|L_i) P_\rmc(M_2|L_j)  p^{2h}_{hs}(v|M_2,M_1,L_j, L_i)\\
+P_\rmc(M_1|L_i) P_\rmnc(M_2|L_j)  p^{2h}_{hs}(v|M_1,M_2,L_i,L_j)\\
+\alpha_i \alpha_j P_\rmnc(M_1|L_i) P_\rmnc(M_2|L_j)  p^{2h}_{ss}(v|M_1,M_2) \big)\\
\times (1+\xi_{12}(M_1,M_2)) N_i N_j 4 \pi r^2 
\end{multline}


\begin{figure*}
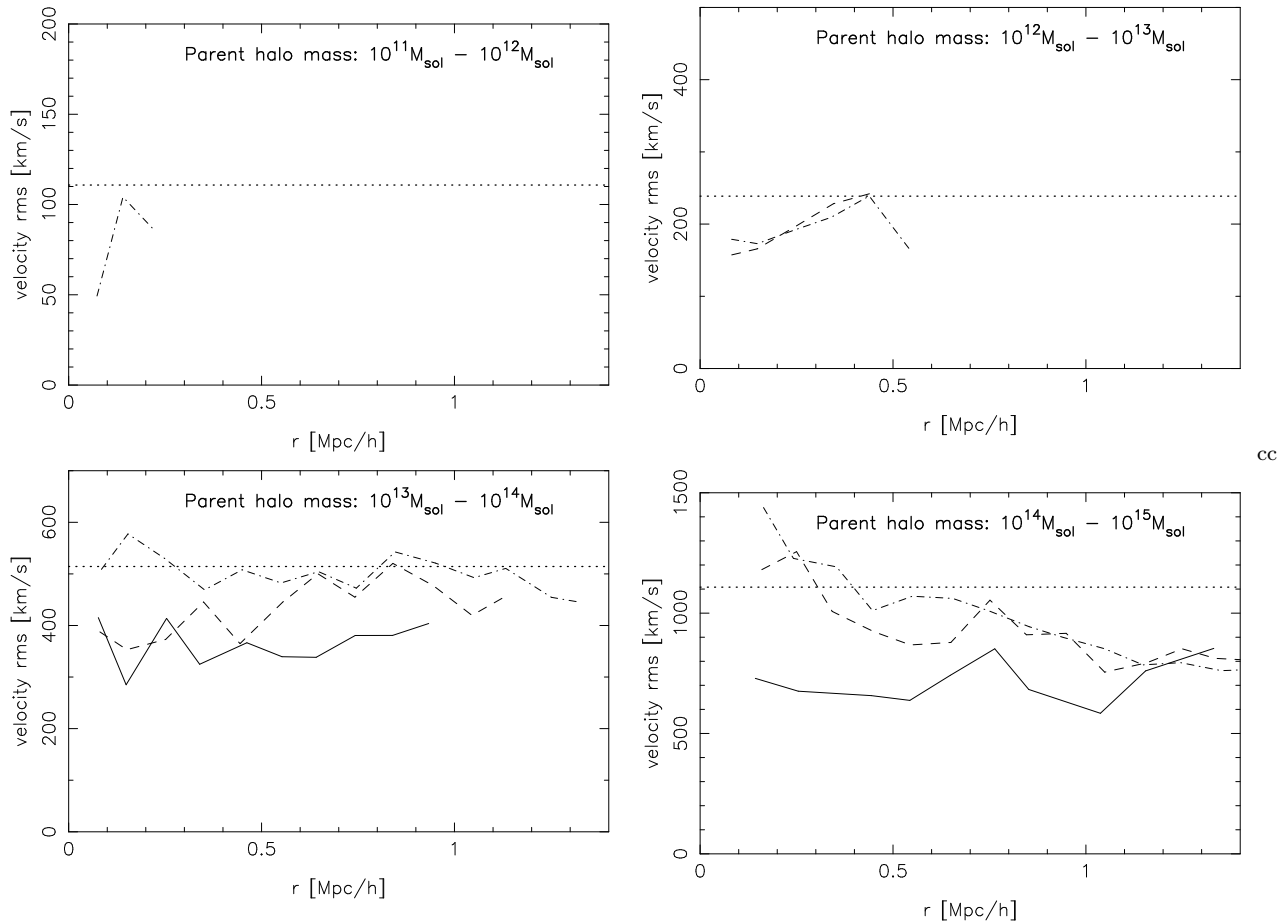

  \begin{tabular}{cc}
  \epsfig{file=c2_11.ps, height=0.45\linewidth, angle=-90} &
  \epsfig{file=c2_12.ps, height=0.45\linewidth, angle=-90} \\
  \epsfig{file=c2_13.ps, height=0.45\linewidth, angle=-90} &
  \epsfig{file=c2_14.ps, height=0.45\linewidth, angle=-90} \\
  \end{tabular}{cc}
  \caption{This figure shows the variation of the one-point velocity
  dispersion relative to the centre of the halo for galaxies belonging
  to a certain parent halo mass range. Line-styles are same as in the
  previous figures. The horizontal dotted line is the virial velocity
  dispersion of the mass corresponding the centre of the (log) mass
  bin. See text for discussion. Note that vertical axes have different
  ranges.\label{fig:vofm}}
\end{figure*}


\section{Basic predictions and comparison with simulations}
\label{sec:basic-pred-comp}

To check the validity of this model we compare it to a high-resolution
numerical simulation. We use collision-less (dark matter only)
simulation that is described in detail in \cite{2004ApJ...614..533T},
which has already been used for comparison with the halo model
\citep{2004astro.ph.10711M}. We used the largest available simulation
with the box size of $120 \mpch$ and $512^3$ particles and the
cosmological model parameters were the same as described in the
previous paragraph. The resolution of the simulation limits the
analysis to halos heavier than $10^{11} \hmodot$ and thus we limit the
discussion to galaxies that are brighter than $-19$ in the r-band
magnitude.

The halos were identified using a variant of the Bound Density Maxima
halo finder \citep{1999ApJ...516..530K}; the details of this algorithm
and the corresponding parameters can be found in
\cite{2004ApJ...609...35K}. A catalog of dark matter halos is
constructed so that every dark matter halo can have zero or more
subhalos. Every halo is assumed to host a central galaxy.  The maximum circular
velocity is used as a proxy for halo mass for both halos and
subhalos. Then the luminosities are assigned to each halo by matching
the cumulative velocity function $n(>V_{\rm max})$ to the observed
r-band cumulative luminosity function \citep{2003ApJ...592..819B}. Since
the mean SDSS redshift is $\sim 0.1$ the same redshift was used in
simulation. More details can be found in \cite{2004ApJ...614..533T}.

\subsection{One-point velocity dispersion}

In Figure \ref{fig:pp1} we show the one-point velocity dispersion
relative to the centre of the halo. This is calculated using the
equation (\ref{eq:10}), but using the one-point velocity distribution
given by the equation (\ref{eq:5}). In other words it is the root mean
square velocity of satellites at distance $r$ from the centre of
their halo. Note that the so-defined one-point velocity dispersion is
different from host-sat contribution to the pairwise velocity
distribution: the latter additionally requires the luminosity of the central galaxy to belong
to a chosen bin.

The velocity dispersion increases with increasing radius, even if we
assume that the velocity dispersions is constant throughout a halo.
This is simply a consequence of halo density weighting.  The one-point
velocity dispersion shows the expected behaviour of increasing
velocity with the increasing luminosity at small separations. For
separations from the centre of the halo that are larger than cut-off
radius of halos of mass $M_{0,i}$ (for the luminosity bin $L_i$), the
velocity dispersion increases faster with radius. At large enough
separations all curves converge to the same one, following the
distribution of velocities in halos weighted $M\eta$, where only halos
with large enough cut-off radius can contribute.

Figure \ref{fig:spp1} shows the one-point velocity dispersion for
galaxies in our simulation.  We note that it does not show the
expected behaviour. The more luminous galaxies live in heavier halos
and thus they should be moving faster. This is a fairly robust
prediction of the theory. The simulation shows that galaxies at the
same distance from the centre move at an approximately constant speed
at all radii, regardless of their luminosity. To understand this behaviour better we
plot the one-point velocity dispersion relative to the centre of the
halo in bins of the parent halo mass in Figure \ref{fig:vofm}. We
split the entire galaxy catalogue into four logarithmic bins according
to the parent halo mass. We then split galaxies in each bin into three
luminosity bins and plot the one-point velocity dispersion, where each
individual point was plotted only if more than five galaxies
contributed to it. We note the following: first, the approximation
that the velocity dispersion is constant throughout the halo is
roughly valid, in fact the velocity drops somewhat with distance as
expected for an isotropic tracer in an NFW profile.  This can also be
a result of increased subhalo bias towards the centre \citep[see
e.g.,][]{2004MNRAS.352..535D}.  Second, the less luminous galaxies
move faster than the more luminous galaxies even in halos of the same
mass. This indicates a velocity bias that is increasing with
decreasing luminosity. It is this bias that in Figure \ref{fig:spp1}
conspires to nearly cancel the fact that one average more luminous
galaxies come from more massive halos.

\subsection{Pairwise dispersion}

\begin{figure}
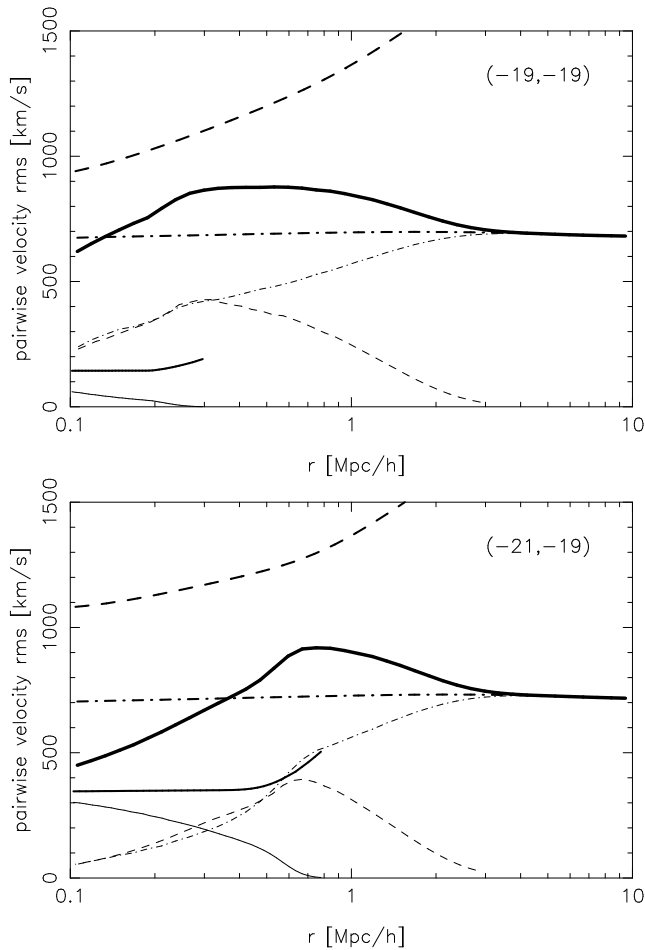

  \epsfig{file=pplot12_19.ps, height=\linewidth, angle=-90} \\
  \epsfig{file=pplot12_21.ps, height=\linewidth, angle=-90}

  \caption{In this figure we show different contributions to the
  pairwise velocity dispersion. The very thick solid line is the total
  velocity dispersion, composed of host-sat pairs (thick solid),
  sat-sat pairs (thick dashed) and two-halo contribution (thick
  dot-dashed).  These
  three contributions are weighted by the respective number of
  pairs. The thin lines show velocity dispersion multiplied by a
  fraction of a given component (so that thin lines add-up to the
  thick line). We plot the ($-19$,$-19$) (top) and ($-21$,$-19$) (bottom) pairs. See
  text for discussion. \label{fig:newfig}}

\end{figure}

We first investigate how individual components contribute to the
pairwise velocity dispersion. We plot these in Figure
\ref{fig:newfig}.  Thick lines are velocity dispersions of different
components. The total velocity dispersion is the weighted combination
of these individual dispersions. To ease visualisation we also plot
individual components multiplied by the fraction of pairs of a given type.

At separations larger than $\sim 2 \mpch$, the two-halo contribution
dominates. At smaller scales there is no regime where one type would
dominate. The ($-21$,$-19$) bin has a somewhat bigger contribution
from the host-sat galaxy pairs, but not dominant enough to allow a clear
separation by distance. The sat-sat contribution has considerably
higher dispersion than the host-sat contribution. These pairs come
from massive halos that have many light satellites and since
dispersion is a pair weighted quantity, these dominate.

\begin{figure}
\epsfig{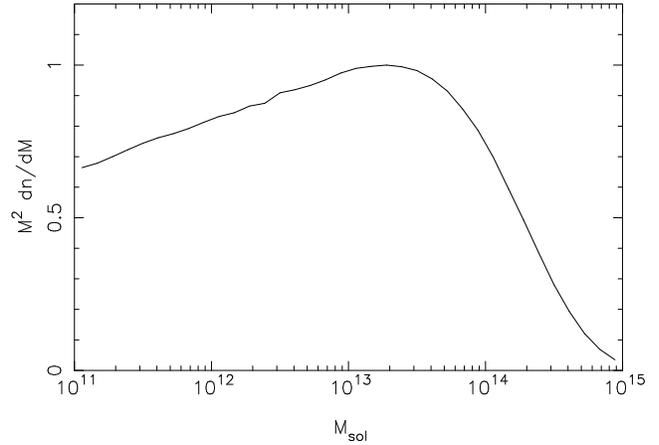}
  
\caption{This figure shows $\eta M^2$ as a function of $M$ and shows
  which mass range is picked up by the sat-sat pairs in our model. See
  text for discussion. \label{fig:etamsq} }
\end{figure}

In Figure \ref{fig:allpp23} we plot the combined contribution to the
pairwise velocities as predicted by our model. At distances below
$1\mpch$ we note that the galaxies selected from less luminous bins
appear to move faster. This is explained by the fact that the fraction
of host-sat pairs is decreasing as we go towards less luminous
galaxies and sat-sat signal is almost independent of luminosity.  This
happens when the lower limit halo mass of satellites is low enough
(i.e., the luminosity of galaxies involved is small enough) that the
dominating sat-sat contribution, which probes $M^{2} \Phi_2(r|M) \eta$
(see Equation~(\ref{eq:5})), approaches a constant.  In Figure
\ref{fig:etamsq} we plot $M^2\eta$ showing that sat-sat pairs
effectively pick up masses around $10^{14} \hmodot$ (though this
of course depends on the power-law index in equation (\ref{nm})). The $\Phi_2(r|M)$
factor additionally biases towards higher masses with increasing $r$.

When the two contributions are combined they result in a pronounced
peak at around the typical size of the largest and therefore the most
massive halos present. Such halos contain a lot of sat-sat pairs that
move very fast and those dominate that region (at around 1
$\mpch$). At larger distance we are dominated by many lighter halos
with only a few members.

\begin{figure}
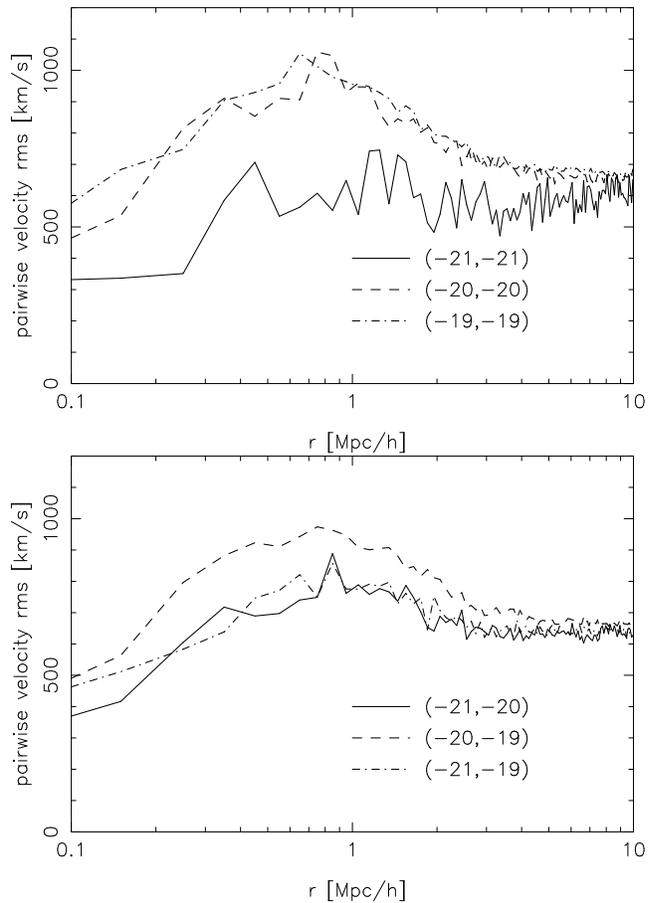

  \epsfig{file=pplot402.ps, height=\linewidth, angle=-90}
  \epsfig{file=pplot403.ps, height=\linewidth, angle=-90}

  \caption{This figure shows the total pairwise velocity dispersion as
  measured in the numerical simulation.  \label{fig:simpp23}}
\end{figure}

In Figure \ref{fig:simpp23} we show the same total pairwise velocity
dispersion as measured in our simulation. We note that the peak
predicted by our simple model is also observed in the simulation.

To summarise the conclusions from this section, we have found that the
pairwise velocity dispersion consists of 3 components, host-sat,
sat-sat and 2-halo pairs and that there is no regime where one
component would clearly dominate over the others.

The sat-sat component has the largest amplitude and drives the
pairwise velocity dispersion peaks at around 1$\mpch$. The amplitude
at the peak is fairly constant and determined by the second moment of
the mass function, which is dominated by large clusters. Pairwise
velocity dispersion cannot be used to determine the typical halo mass
of galaxies except on very small scales, where the number of pairs is
small.  This is true even for cross-correlation between different
luminosity samples, which in principle is a better way to select
host-sat pairs: while this does increase the host-sat to sat-sat ratio
we find that it still does not sufficiently suppress the number of
sat-sat pairs to make it a reliable tracer of host dark matter
halo. Thus additional selections based on galaxy environment are needed
if one wants to select a host-satellite sample only.  Such additional
selections may be difficult to model and reduce the size of the
sample.

\subsection{Adding velocity bias}
\label{sec:adding-velocity-bias}

We are now in position to rectify the main deficiency of our model
with respect to the simulation, namely the velocity dependent bias.
This was achieved by using the following values of the velocity bias
as defined in Equation \eqref{eq:8}:
$b_{\rm v} (-19)=1.0$, $b_{\rm v} (-20)=0.9$, $b_{\rm v} (-21)=0.6$.
These values were chosen so that the model
reproduced the correct behaviour of the pairwise velocity dispersion.
Results are plotted in Figure \ref{fig:plus} (note that individual
lines were shifted vertically by the same amount to make plot
readable). The fit is remarkably good considering the simplicity of
our model.

\subsection{Fraction of satellites}
\label{sec:fraction-satellites}

\begin{figure}
  \epsfig{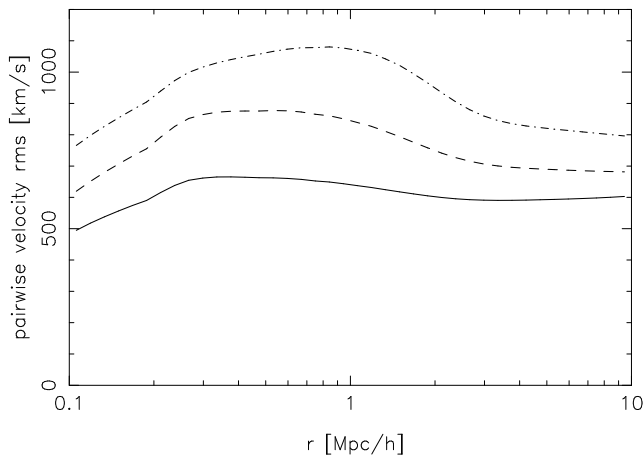}

  \caption{ In this figure we plot the effect of changing the
    satellite fraction on the velocity dispersion. All lines correspond
    to ($-19$,$-19$) pairs the satellite fraction $\alpha$ set to 0.4
    (dot-dashed), 0.2 (dashed) and 0.1 (solid). Increasing the
    $\alpha$ parameter increases the overall velocity dispersion, but
    also makes the $1\mpch$ bump more pronounced.
 \label{fig:alpha}}

\end{figure}

A similar effect to the effect of velocity bias can be achieved by
assuming a luminosity dependent fraction of satellites, i.e.,
$\alpha=\alpha(L)$.  As Figure \ref{fig:alpha} shows, increasing the satellite
fraction leads to an increase in velocity dispersion, as expected.  It
also makes the peak at $r=1\mpch$ more pronounced, since that peak is
caused by satellite-satellite pairs.

The numbers for the fraction of satellites in
our simulation are 0.15, 0.20 and 0.23 for luminosity bins $-19$, $-20$ and
$-21$ respectively. Without introducing velocity bias it is possible to
match the velocity dispersion of  single-luminosity bins as shown
in Figure \ref{fig:plusalpha}. In this Figure we used the values
as low as 0.06 for the most luminous galaxies, 0.17 for $-20$ galaxies
and 0.20 for faintest galaxies considered. 

However, as Figure \ref{fig:plusalpha} shows, the values of $\alpha$ that produce
a decent fit for galaxy pairs of a single luminosity bin considerably
decrease the goodness of fit for cross-bins. We also note that
decreasing the fraction of satellites decreases the height of the peak at
around $r\sim 1\mpch$.


\begin{figure}
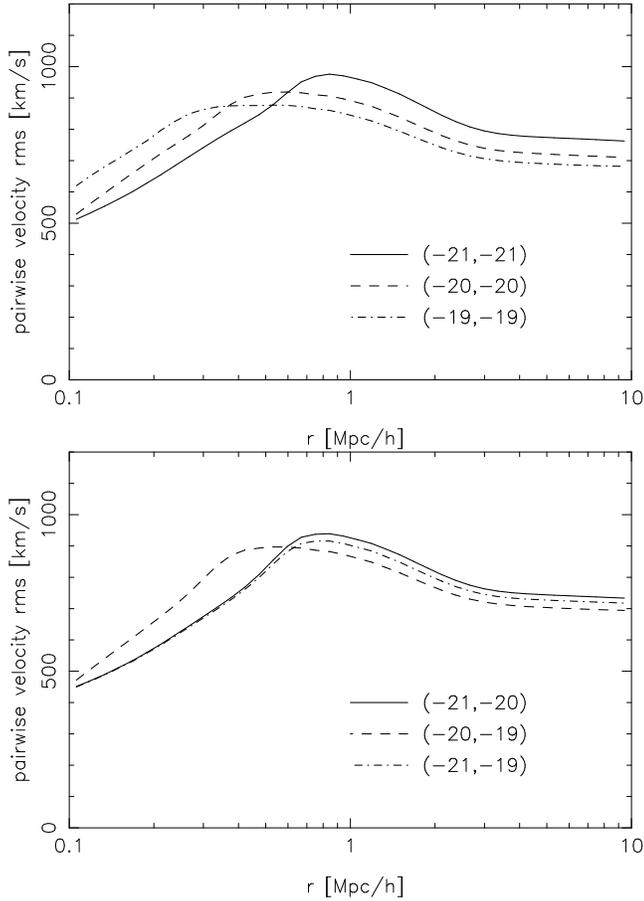

  \epsfig{file=pplot202.ps, height=\linewidth, angle=-90}
  \epsfig{file=pplot203.ps, height=\linewidth, angle=-90}

  \caption{This figure shows the total prediction for the pairwise
  velocity dispersion for various luminosity bins shown in the
  legend. Distances below 1 $\mpch$ are dominated by the one-halo
  term, while the  two halo term dominates at larger
  distances. \label{fig:allpp23} }
\end{figure}

\begin{figure}
  \epsfig{file=pplot+202.ps, height=\linewidth, angle=-90}
  \epsfig{file=pplot+203.ps, height=\linewidth, angle=-90}

  \caption{This figure shows the total pairwise velocity dispersion as
  measured in the numerical simulation compared to our model when
  velocity dependent bias has been added to the model. Jagged thin
  line are simulation data and smooth thick line the model
  prediction. Lines have been shifted vertically to ease
  visualisation. \label{fig:plus}}

\end{figure}

\begin{figure}
  \epsfig{file=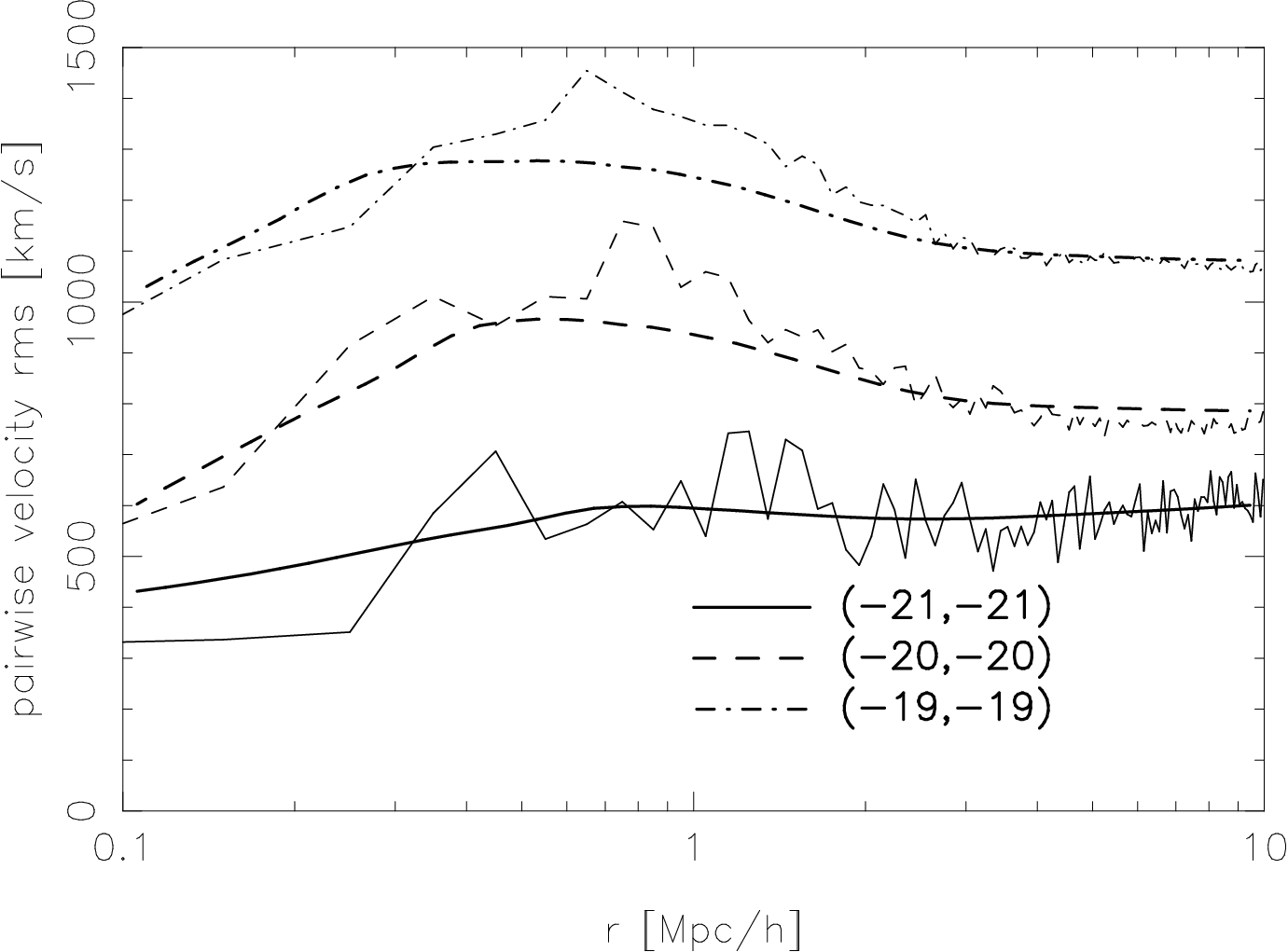, height=\linewidth, angle=-90}
  \epsfig{file=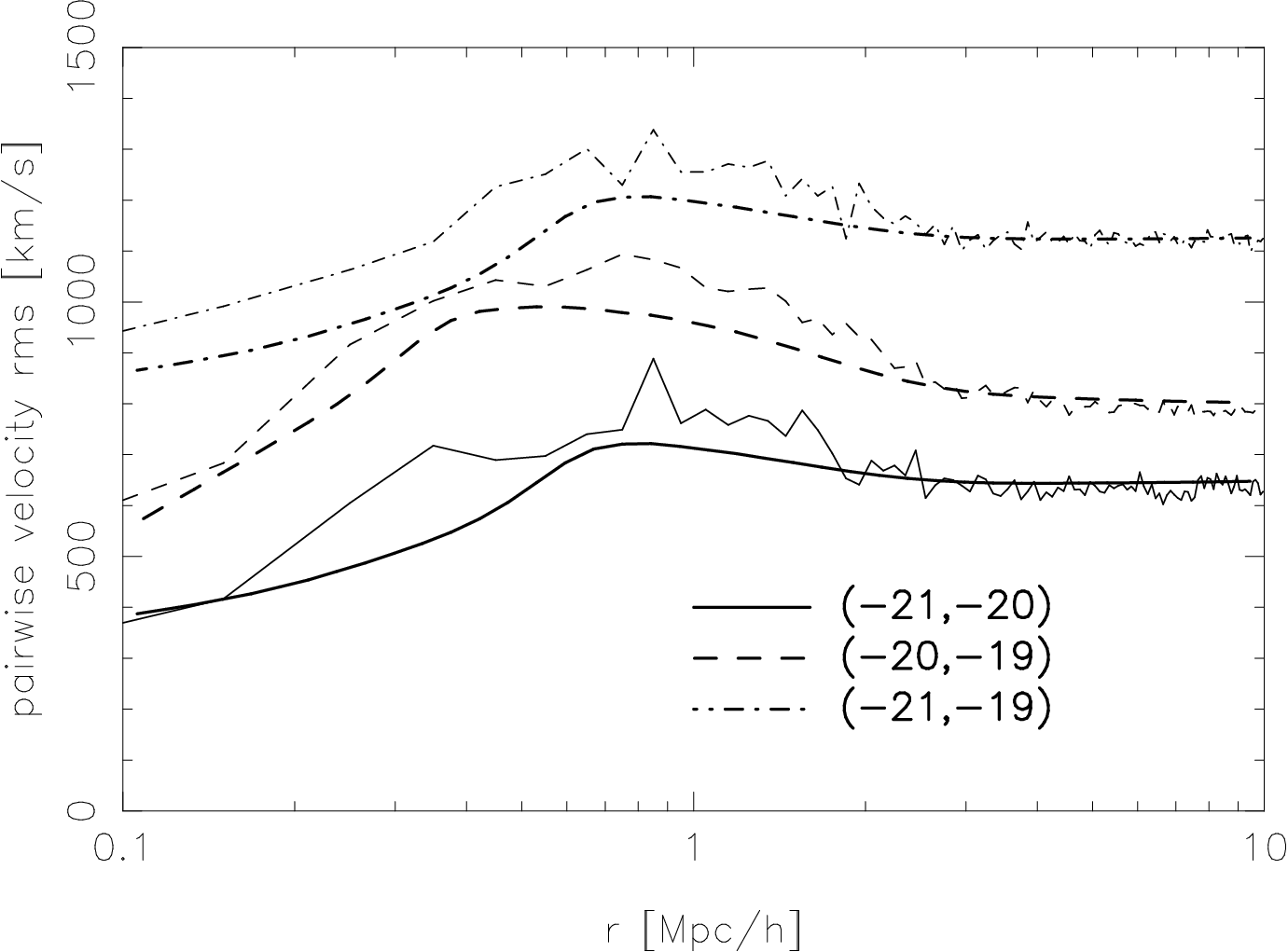, height=\linewidth, angle=-90}

  \caption{ Same as Figure \ref{fig:plus}, but with variable fraction
  of satellites used to match single-luminosity bins, instead of
  luminosity dependent velocity bias. See text for discussion.  
 \label{fig:plusalpha}}

\end{figure}


\section{The observable correlation function}

\label{sec:distr-r_pr-pi}

\subsection{Correlation function  and the halo model}

The pairwise velocity distribution we have been calculating so far is
a basic distribution from which various other statistics can be
calculated. It is not directly observable by itself. In the flat sky
approximation, we see objects at various projected radii and their
redshift encodes both their distance and their velocity component.

At large separations corresponding to two-halo contributions the
galaxies experience infall that is well described by the linear theory
beta model of \cite{1987MNRAS.227....1K}. This results in squashing of
the correlation function along the line of sight. The amount of
distortion is parametrised by a single parameter $\beta=f/b$, where
$f\sim \Omega_m^{0.6}$ is the linear growth factor and $b$ is the linear
bias. Ideally, one would like to incorporate the infall by considering
a distribution of infall velocities. Note that
\cite{2004PhRvD..70h3007S} has shown that no distribution of infall
velocities reproduces the beta model. Nevertheless, since beta model
appears to be a good phenomenological model for description of infall,
we include it here.

First we rewrite the probability of finding an object between $r$ and
$r+\rmd r$ and having velocity $v$ between $v$ and $v+\rmd v$ into
cylindrical coordinates $r_\proj$ and $z$:

\begin{equation}
  P_{\cyl} (r_\proj, z,v) = P\left(\sqrt{r_\proj^2+z^2},v\right) \frac{r_\proj}{2\left(r_\proj^2+z^2\right)}
\end{equation}

The distribution in the $r_\proj - \pi$ plane is then given by

\begin{equation}
  P_{\rm obs} (r_\proj, \pi) = \int P_\cyl^{\rm 1D} \left(r_\proj,
  z,H_0(\pi-z)\right) \frac{\rmd z}{H_0}.
\label{eq:11}
\end{equation}

We then calculate the observable correlation function by summing up
contributions from 1-halo and 2-halo terms, following a procedure
similar to many other works \citep{2001MNRAS.321....1W,2001MNRAS.325.1359S,2002MNRAS.336..892K,2004MNRAS.348..250C}:

\begin{equation}
  \xi (r_\proj,\pi) = \xi^{1h}(r_\proj,\pi) + \xi^{2h} (r_\proj,\pi)
\end{equation}

The one-halo contribution is calculated from the Equation
(\ref{eq:11}). This implicitly assumes that these systems are
completely virialised and therefore not affected by the infall. The
two-halo contribution, on the other hand, is calculated using standard
dispersion model, which takes the following simple form in Fourier space:
\begin{equation}
P^S(\vect{k})=P^R(k)\frac{(1+\beta \mu^2)^2}{1+k^2\mu^2\sigma_{12}^2/2}.
\label{eq:12}
\end{equation}
Here $P^S$ and $P^R$ denote the power spectrum in redshift space and
real space respectively and $\mu$ is angle to the line of sight.  The
$\beta$ parameter is calculated using bias predicted for a given
luminosity bin from our model and velocity dispersion is given by the
equation (\ref{eq:10}).

\subsection{Comparison to observations}

\cite{2004ApJ...617..782J} have found that less
luminous galaxies in the 2dF redshift survey catalog appear to move
faster. This conclusion is reached as follows. The correlation
function of galaxies in the 2dF catalog is calculated for galaxies
selected from various luminosity bins.  This correlation function is
used to calculate the redshift-space power spectrum
$P^{S}(\vect{k})$. For each value of $k$ the data is fit by fixing
$\beta=0.45$ and fitting for $\sigma_{12}$ and $P(k)$.

\begin{figure}
  \epsfig{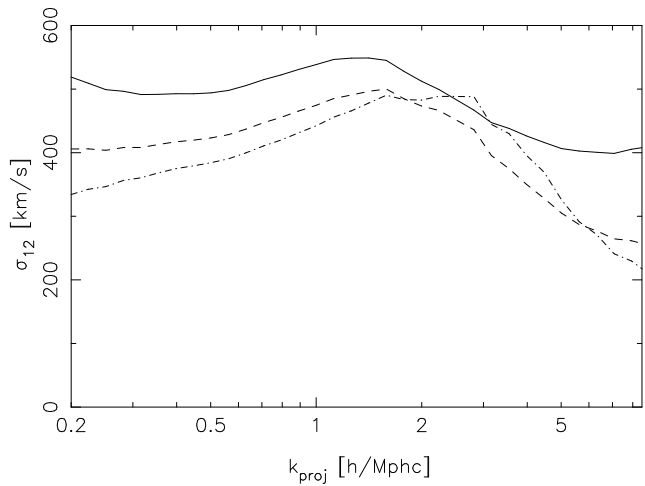}
  \caption{This figure shows the recovered $\sigma_{12}$ parameter
following the prescription used to analyse the data in
\citep{2004ApJ...617..782J}. Line styles as before.}
\label{fig:jbpf} 
\end{figure}

There are three features in our model that can lead to an explanation
of this effect. First, it is possible that the effect of pair
weighting gives enough biasing towards more massive systems to explain
the effect. As shown in Figure \ref{fig:jbpf} the recovered velocity dispersion 
is non-monotonic, allowing for the possibility that velocity dispersion 
decreases with luminosity over some range, although  
the effect is not very pronounced and depends on the scale. 
Second, the addition of the luminosity dependent
velocity bias works further in making less luminous galaxies appear to
move faster. And third, allowing for satellite fraction to 
decrease with luminosity also leads to a decrease of velocity dispersion. 

In order to meaningfully compare our predictions with Jing \& B\"orner
result we mimic their procedure. We predict a redshift space
correlation function using our model, calculate the power spectrum and
then fit the dispersion model to it. We also follow their procedure by
fixing the $\beta$ parameter to 0.45.

\begin{figure}
  \epsfig{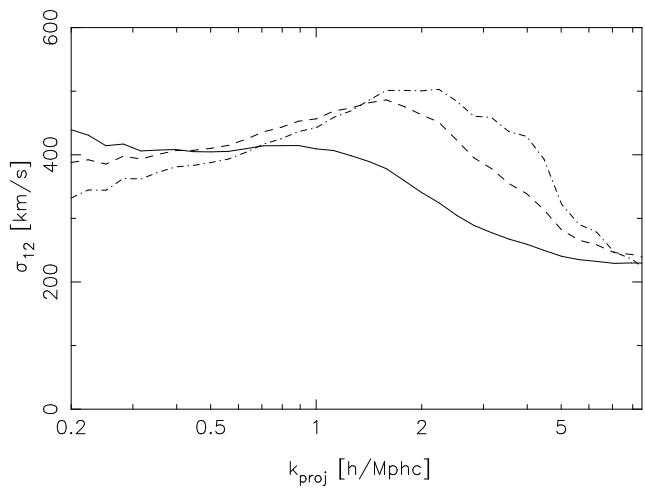}

  \caption{Same as figure \ref{fig:jbpf}, but a luminosity
    dependent velocity dispersion has been added to the model.}  
\label{fig:jbpf2}
\end{figure}

The main results of this exercise are plotted in Figures
\ref{fig:jbpf} and \ref{fig:jbpf2}. We plot the recovered value of the
$\sigma_{12}$ parameter as a function of the wave vector $k$ for three
different luminosity bins. Figure \ref{fig:jbpf} shows the behaviour
for the simple model discussed throughout this paper. We note that the
plot looks qualitatively the same as Figure \ref{fig:allpp23} with an
inverted horizontal axis. Here too, there is a typical bump at $k \sim
1 h / {\rm Mpc}$.  The pair-weighting can result in less luminous
galaxies appearing to move faster, but only at $k \gtrsim 2 h/{\rm Mpc}$ or
so and therfore it is not clear whether this effect alone provides
satisfactory explanation.

In Figure \ref{fig:jbpf2} we repeated the analysis, but added a
luminosity dependent velocity bias that was used in section
\ref{sec:adding-velocity-bias}. Velocity biasing has the expected
effect on the inferred velocity dispersions: galaxies of lower
luminosity can indeed have a lower $\sigma_{12}$ even at wave vectors
as large as $1 h/{\rm Mpc}$.

We note that a similar analysis of the SDSS data
\citep{2005astro.ph..9874L} shows qualitatively similar behaviour to
our predictions plotted in Figure \ref{fig:jbpf2}: a wide bump at
around $1-2 h/{\rm Mpc}$ that becomes less pronounced for higher
luminosities. The most luminous bins in that paper show an increase of
$\sigma_{12}$ with increasing $k$. We do not observe this behaviour
here.

\section{Discussion and Conclusions}
\label{sec:conclusions}

In this paper we have investigated the pairwise velocity dispersion in
the halo model and compared it to simulations.  The distribution of
velocities for galaxies separated by a certain distance in the real
space is a pair weighted quantity and thus does not reflect the
velocity structure of halos typically associated with galaxies of a
given luminosity. In particular, for the pairs selected from either
unequal luminosity bins or from equal luminosity bins
(e.g., ($-20$,$-19$), ($-19$,$-19$), etc.) the main contribution comes
from the sat-sat pairs.  Since the halo mass distribution of
satellites follows the halo mass function, which is exponentially
suppressed on high mass end and has a power law at the low mass end,
it probes a reasonably restricted range of parent halo mass around
$10^{14} \hmodot$ and is nearly a constant. This, in combination with
the fact that if one has a larger number of galaxies the number of
sat-sat pairs relative to host-sat pairs increases, results in
velocity dispersion which is not very luminosity dependent. The pairs
selected so that one galaxy is more luminous that the other are also
dominated by sat-sat pairs.
 
We have made predictions for redshift space correlation function of
galaxies and fit it with the dispersion model, mimicking the
observation procedure employed in \cite{2004ApJ...617..782J} and in
\cite{2005astro.ph..9874L}.  In the latter work the measured
$\sigma_{12}$ in SDSS data has a qualitatively similar dependence on
the wavenumber as predicted by our model.  One of the main conclusions
of these papers is that velocity dispersion declines with luminosity,
which appears to contradict simple halo models.  As mentioned even the
simplest models do not suggest that there is a strong luminosity
dependence, since most of the contribution comes from sat-sat pairs.
Two other effects can give rise to velocity dispersion decreasing with
luminosity. One is luminosity dependent velocity bias.  In the
simulation the effect that more luminous galaxies appear to move
slower is clearly evident: in the halos of the same mass, the more
luminous subhalos tend to move slower than their less luminous
counterparts.  Tidal stripping and tidal disruption have been
suggested to explain the anti-bias of subhalo distribution \citep[see
e.g.,][]{1999ApJ...520..437K,2004MNRAS.352..535D,2005ApJ...624..505Z}
to the dark-matter distribution in a given halo and to explain the
velocity bias of the subhalo population. Subhalos with lower orbital
energy do not survive and one is left with halos that are on average
faster. We note however that in these dark matter simulations one uses
a simple prescription to link the luminosity of a galaxy with the
circular velocity of the subhalo, which may be more complicated in the
real world.  Second effect that can give rise to velocity dispersion
decreasing with luminosity is if satellite fraction also decreases
with it. There is already some observational evidence for this from
halo modelling of luminosity dependent galaxy correlation function
\cite{2005ApJ...630....1Z} and this explanation has also been put forth by
\cite{2004ApJ...617..782J}.  Both of these effects can therefore
explain the observed trends and can be included in halo models. In
summary, using auto and cross correlations between luminosity bins in
redshift space to extract PVD is unlikely to provide us with simple
information such as the mass and mass profile of the central galaxy
halos, but is instead telling us more about the fraction of galaxies
that are satellites and their velocity bias inside clusters.

\section*{Acknowledgements}
We acknowledge useful discussions with Andrey Kravtsov.  AS thanks
Rachel Mandelbaum for helping with SDSS and simulation data. AS is
supported by the Slovene Ministry of Higher Education, Science and
Technology.  US is supported by a fellowship from the David and Lucile
Packard Foundation, NASA grants NAG5-1993, NASA NAG5-11489 and NSF
grant CAREER-0132953. AT is supported by the National Science
Foundation (NSF) under grants No.  AST-0206216 and AST-0239759, by
NASA through grants NAG5-13274 and NAG5-12326, and by the Kavli
Institute for Cosmological Physics at the University of Chicago.

\label{lastpage}
\bibliography{../BibTeX/cosmo,../BibTeX/cosmo_preprints}
\bibliographystyle{mnras}
\bsp

\end{document}